\documentclass[manuscript]{acmart}
\AtBeginDocument{%
  \providecommand\BibTeX{{%
    \normalfont B\kern-0.5em{\scshape i\kern-0.25em b}\kern-0.8em\TeX}}}

\usepackage{xcolor}
\usepackage{soul}
\usepackage{hyperref}
\usepackage{booktabs}
\usepackage{multirow}
\usepackage{colortbl}
\usepackage{tikz}
\usetikzlibrary{calc,positioning}
\usepackage{array}
\usepackage{tabularx}
\usepackage[commandnameprefix=always]{changes}
\usepackage{mathptmx}                  
\usepackage{balance}                   
\usepackage[capitalise]{cleveref}

\newcommand{\mysec}[1]{\smallskip \noindent \textbf{#1:}}
\begin{document}

\title{Nanomentoring: Investigating How Quickly People Can Help People Learn Feature-Rich Software}

\author{Ian Drosos}
\email{ian@trent.ai}
\additionalaffiliation{%
  \institution{Trent AI}
  \city{London}
  \country{United Kingdom}
}
\affiliation{%
  \institution{Autodesk Research}
  \city{Toronto}
  \state{Ontario}
  \country{Canada}
}

\author{Jo Vermeulen}
\email{jo.vermeulen@autodesk.com}
\affiliation{%
  \institution{Autodesk Research}
  \city{Toronto}
  \state{Ontario}
  \country{Canada}
}

\author{George Fitzmaurice}
\email{george.fitzmaurice@autodesk.com}
\affiliation{%
  \institution{Autodesk Research}
  \city{Toronto}
  \state{Ontario}
  \country{Canada}
}

\author{Justin Matejka}
\email{justin.matejka@autodesk.com}
\affiliation{%
  \institution{Autodesk Research}
  \city{Toronto}
  \state{Ontario}
  \country{Canada}
}

\renewcommand{\shortauthors}{Anon}

\begin{abstract}
People frequently use online forums to get help from experts to answer questions about feature-rich software. 
However, they may have to wait minutes, hours, or even days to receive advice. 
We investigate the potential to leverage experts to provide quicker help. 
We collected over 200 questions from online forums for two feature-rich software applications and suspected a quarter were short enough to be answered in less than one minute (defined as \emph{nanoquestions}).
We then conducted a study with 28 experts recruited from help forums to confirm this assumption, and explore whether there was a preference between text and audio answers. 
For more than half of the nanoquestions participants saw, they could give advice that they believed was helpful in under 60 seconds. Finally, we collected feedback about what makes a question quick to answer to inspire the design of future tools for ultra rapid human-to-human help. 
\end{abstract}



\begin{teaserfigure}
    \centering
    \includegraphics[width=.9\textwidth]{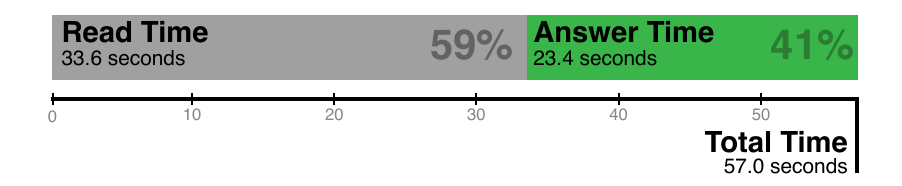}
    \caption{We conducted a study with 28 online forum participants to provide answers to \emph{nanoquestions}, existing forum questions that we suspected could be answered in less than 60 seconds. Participants were often able to give advice very quickly that they believed was helpful. Here, we show the average timeline for the two-phases of answering these questions during our study. Participants spent 58.9\% on average reading the question before engaging with text or audio features for giving advice. The remaining 41.1\% of the time was spent on answering the question.}
    \Description{We conducted a study with 28 online forum participants to provide answers to \emph{nanoquestions}, existing forum questions that we suspected could be answered in less than 60 seconds. Participants were often able to give advice very quickly that they believed was helpful. Here, we show the average timeline for the two-phases of answering these questions during our study. Participants spent 58.9\% on average reading the question before engaging with text or audio features for giving advice. The remaining 41.1\% of the time was spent on answering the question.}
    \label{fig:timeline-teaser}
\end{teaserfigure}


\maketitle

\section{Introduction}
\label{sec:intro}
When users need help with feature-rich software, they frequently use web-based resources to get assistance. One common resource is online forums, which offer access to community experts who answer questions posed by other members. 
However, receiving answers to these questions can take hours, or even days, and sometimes help does not come at all. This holds true for complicated multipart questions, but also for simple queries where a user needs just a little advice to get them unstuck, or receive subjective feedback. 
For example, Hellman et al. reported that the median response time for open-source software forums was 10.83 hours \cite{Hellman2022}. This wait may be fine for some questions, but users who are stuck may need immediate help to continue their task.


Previous work such as MicroMentor~\cite{NikhitaMicromentor2020} has looked at providing quick, synchronous help through short video chat sessions. In this paper, we investigate how quickly help can be provided asynchronously. Specifically, 
we wondered if some questions currently being asked in traditional online forums could instead be provided to, and answered by, experts extremely quickly, without the need for a synchronous one-on-one session. 
While of course not \emph{\textbf{all}} questions are candidates to be answered extremely quickly, \emph{if even a fraction of the questions on forums could be answered in 60 seconds or less, that could lead to a huge productivity gains}.
We define these types of questions as \emph{nanoquestions}, which is any question that can be answered in less than 60 seconds by an expert. 
Nanoquestions are short knowledge gaps or qualitative questions that can help a user get unstuck or get feedback quickly. We want to know if these nanoquestions exist, and what limitations surround them. Based on these findings, future work can address these limitations and support quick-help when nanoquestions are useful.
%

In this work we seek to answer three questions surrounding nanoquestions for getting help quickly:
\vspace{-0.3em}
\begin{itemize}
    \item Can helpers give advice quickly that they believe is helpful?
    \item Do helpers have a preference of modality (text vs. audio) when giving advice quickly?
    \item What challenges do helpers face when giving advice quickly?
\end{itemize}

To discover how to support this quick-help workflow, we performed an exploratory study of nanoquestions and getting help quickly.
First, we extracted 204 questions posted on online forums about two feature-rich applications (Autodesk Fusion~360 and Microsoft Word) and classified them based on their potential to be answered in less than 60 seconds. We labeled these as potential nanoquestions.
We then asked 28 online forum participants to answer 20 of these questions using text or audio, and provide feedback on what would make these questions easier to answer in a short amount of time. Our participants were often able to give answers very quickly that they believed were helpful, had varying modality preferences and perceived strengths and weaknesses of each, and provided feedback on the difficulties of providing help quickly. We also discovered insights for crafting better nanoquestions and future implications for the design of quick-help interactions.


The contributions of this paper are: 
\vspace{-0.1em}
\begin{itemize}
\itemsep0.2em 
    \item An exploratory study of nanoquestions to determine if quick (less than 60 seconds) answers is an achievable goal for real questions found online (\cref{sec:results}).
    \item Design implications and future directions for rapid help systems aimed at feature-rich software (\cref{sec:disc}).
\end{itemize}


\section{Related work}
\label{sec:rw}
\subsection{Systems for Getting and Giving Help}
Several systems have been developed that integrate with applications to provide relevant help through videos or expert advice.
For example, IP-QAT streamlines the process for asking software-related questions by integrating with the application itself \cite{MatejkaIpqat2011}.
RePlay presents users with educational videos based on the application being used and tasks being done \cite{FraserReplay2019}, while ReMap assists in the searching of these videos within the applications  \cite{FrasaerRemap2020}.

Other systems such as LemonAid \cite{ChilanaLemonaid2012} leverage crowdsourcing to provide advice and feedback by using previously asked questions and answers, while Codeon verbally asks questions to many helpers to receive asynchronous answers \cite{ChenCodeon2017}, and Rsourcer obtains specific feedback through ``micro-reviews'' and follow-ups with mentors \cite{JiangRsourcer2023}.
For further related work in the domain of in crowdsourcing for education, Jiang et al. provides a thorough review \cite{JiangReview2018}.

Another collection of systems target the task of finding the experts that are best suited to answer a question. Finding the ``right'' expert is critical for getting an answer for a specific question \cite{McdonaldExpertise2000}. Some approaches involve leveraging a user's own social network of experts to route questions \cite{HorowitzAardvark2010}, leveraging existing answers to commonly asked questions and going to an expert when a new question is asked \cite{AckermanAnswergarden1990, AckermanAnswergarden21996}, or building profiles for experts based on questions they had previously answered to help route questions \cite{RiahiFindingexperts2012}. 


Finally, there are several systems for supporting real-time collaboration between learners for cooperative learning \cite{AlharthiTwoTorials2022} or between learners and experts for tutoring \cite{GuoCodechella2015}.
MarmalAid, a 3D modeling system for novices, allows users to receive real-time help from experts and easily provide question context by sharing a link to the document being worked on that shows what part of the model the novice is asking about \cite{ChilanaRealtimeexperts2018}. This provides context to the expert, so they can better understand what the learner is asking about.
To speed up getting mentorship from experts, MicroMentor connects novices with experts for 3-minute live one-on-one help sessions over a video call, which was found to be a sufficient amount of time to provide live help in many situations \cite{NikhitaMicromentor2020}.

Our work investigates if users can get help and mentorship from experts \emph{even faster} than the quick help provided by existing systems (e.g., MicroMentor's 3-minute sessions), and identify friction points which may prevent ultra-quick help from being possible.

\subsection{Online Communities and Help}
Other research has investigated help communities and the best ways to support askers in obtaining assistance from experts.
One study looked at barriers preventing people from answering coding questions, which include feeling like their expertise is not enough, and time constraints \cite{FordBarriers2016}.
Ford et al. also found that mentorship from experienced community members assisted novices in forming higher quality questions (e.g., better titles, context) to online question-and-answer communities and noted there were advantages to human-to-human mentorship (e.g., increased participation from novices, overcoming language barriers) \cite{FordCollabEditing2018}. 
Audio-based collaborative sessions have been shown to help in supporting mentorship
\cite{VarghesePeerMentoring2022}, as well as video-based collaborative sessions where
livestreamers provide expert knowledge to novices by answering chat questions, showing tacit knowledge remains a challenge \cite{DrososStreamersTeaching2021}.

Relating to challenges to asking questions and getting help, Kiani et al. found that 
novices can face challenges with expressing their issues and struggle in recognizing when a help resource is relevant \cite{KianiBeyondonesize2019}. Similarly, 
Biehl et al. found that common ground between the asker and the helper was needed to properly address a question and noted a need for tools that help the asker express the issue they are facing  \cite{BiehlHelpDesk2022}.
Supporting our goal of getting users help in less than 60 seconds, Kotturi et al. found responsive support can be critical in this asker-helper paradigm
\cite{KotturiTechHelpDesk2022}.

To explore how to design better and easier to answer questions, a study found question clarity and effort to answer play a part in why some Stack Overflow questions go unanswered; however, attracting an expert is still an issue even if the question is well-formed \cite{AsaduzzamanAnswering2013}. 
Other work also found lack of clarity as a main issue preventing the answering of questions asked online, and found that a lack of screenshots or videos of the issue contributed to potential helpers needing more information before being able to assist \cite{HudsonUnderstanding2015, ChilanaMultimedia2011}. 
In a study of experts providing advice to web programming questions, helpers commonly wanted more context to the asker's question while askers needed a quick and immediate response \cite{ChenOnDemand2016}, which multimedia may provide over text. 
Learners frequently attempted to minimize the impact of the time taken from experts providing mentorship through several strategies like adding context to the question and switching to another task while waiting for assistance~\cite{BerlinConsultants1992}.

Our work builds on this research to understand what requirements are needed in a question so that a helper is able to very quickly provide an answer or advice.

\section{Study Methodology}
\label{sec:methods}
\subsection{Tasks}

We aimed to discover how feature-rich software learners could get help from experts, and specifically, how they could get help more quickly than in on online forums, or even faster than the 3 minutes afforded by MicroMentor \cite{NikhitaMicromentor2020}. 

First, we searched through various online resources for getting help for the tools Autodesk Fusion~360 and Microsoft Word. We leveraged official forums \cite{FusionOfficial, WordOfficial} 
and Subreddits \cite{FusionReddit, WordReddit} 
dedicated to each application to find real questions being asked about complex software. While of course not \emph{all} questions being asked in online forums can be answered in less than one minute (which we define as a \emph{nanoquestion}), we speculate that at least some of them can be. 
The lead author collected 204 forum questions with responses, split between Word and Fusion~360. These were the most recent active questions at the time. Questions without responses were skipped, so answers could help determine if a question was a potential nanoquestion.
Two researchers judged each question's potential to be answered in less than 60 seconds by selecting questions which were shorter, clear, and did not contain lengthy videos.
This resulted in 30 questions for Fusion~360 and 38 questions for Word as candidate nanoquestions. 
While sampling from the entire history of forum posts may be more generally representative, 
we wanted to capture a snapshot of user questions being addressed by experts. However, as software changes, so may the questions.

Questions ranged from UI navigation (`Where is 'Import File''') to qualitative feedback (``Is this [model of tank] real enough?'').
The average body length of the selected questions
was 57.59 words and 8.38 words for titles. For context, in data collected from 12 StackExchange subforums~\cite{CQADupStack2015}, the average body length is roughly double that (116.52 words; range: 83.4--166.5).
Based on an average 228 WPM reading speed in English \cite{readtime}, we would expect around a 17-second read time for our questions. Even with viewing images attached to some questions, we believed this afforded experts enough time to read the question, and submit a short text or audio answer in less than a minute.

\subsection{Participants}
Participants were recruited from both official help forums for the software applications and relevant Subreddits. We asked for volunteers with experience helping people online or at work in Autodesk Fusion~360 or Microsoft Word. We recruited 28 participants (8 women, 20 men), 13 for Fusion~360 (2 women, 11 men), and 15 for Microsoft Word (6 women, 9 men), 
selected for their experience in answering questions on forums.
Participants had professions ranging from students to professional engineers (e.g., civil, mechanical, software), accountants, and 3D modelers. Participants self-rated as frequently helping others online (median = ``Weekly'' (3)) and offline (median = ``Daily'' (4)) with their chosen application, using a 5-point Likert scale from ``Never'' (1) to ``Several Times a Day'' (5). 

\subsection{Protocol}
\begin{figure}[ht]
\includegraphics[width=.9\columnwidth]{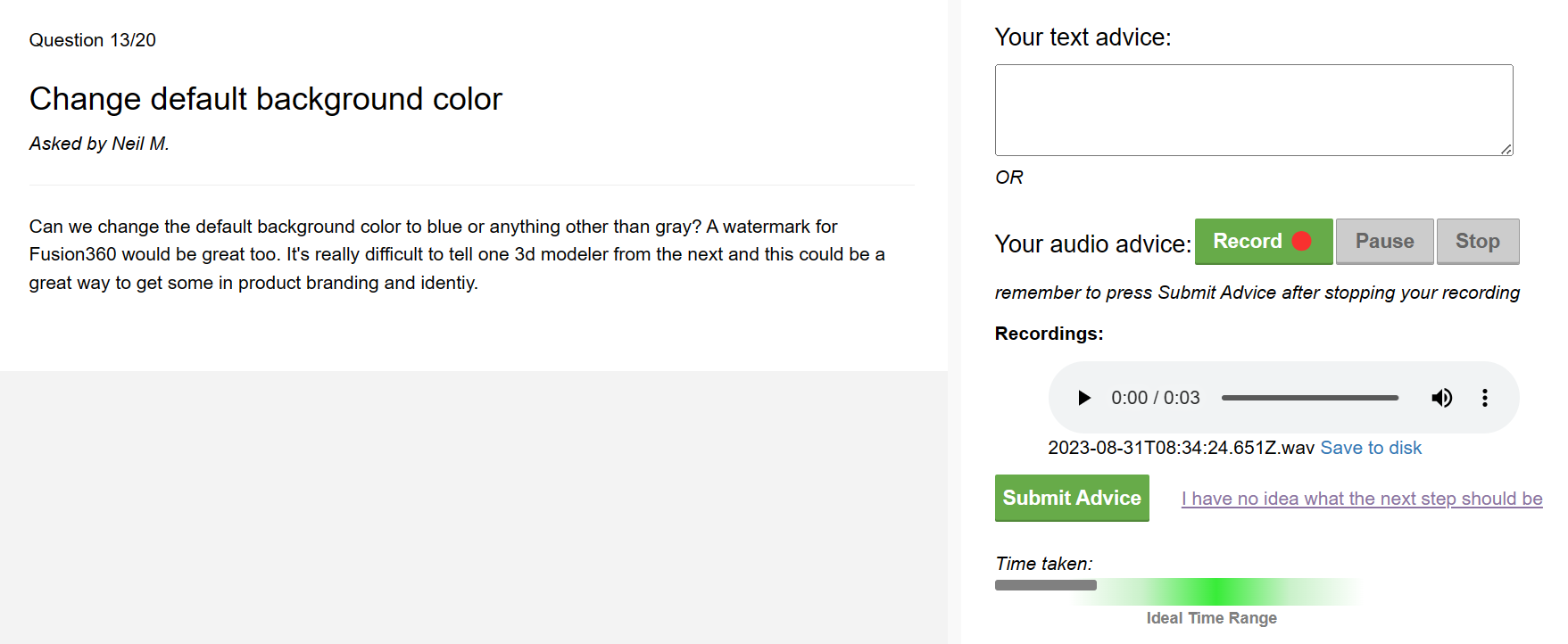}
  \caption{Survey question UI that participants saw. Participants could answer the question based on their preferred modality.}
  \Description{Survey question UI that participants saw. Participants could answer the question based on their preferred modality.}
  \label{fig:surveyimage}
\end{figure}

We developed a custom JavaScript web application to serve our collected nanoquestions to participants (\cref{fig:surveyimage}). Participants received 20 questions relating to the application they were familiar with in random order. The system allowed participants to answer the questions with both text and audio. The modality in which participants provided their advice was counterbalanced. For the first five questions, half of the participants used text, while the other half used audio recordings. For the next five questions, participants used the opposite modality. Finally, the last 10 questions allowed the participant to choose which modality to provide their answers, based on their personal preference. 

After a tutorial, participants were saw one question at a time and gave advice. These questions had the text (and any images or videos) from original posts formatted for our tool. Participants had a text area, audio recording button, or both depending on the stage of the study they were in and were instructed to read the question and quickly give the best answer they could. A progress bar under the question showed how long they had spent on the question to encourage them to answer quickly (\cref{fig:surveyimage}, bottom right). 
A visual alert warned participants after one minute. 
However, participants could take as long as they needed to submit an answer.

After the participant provided their advice, they pressed a submit button and answered a short survey about that specific question (e.g., ``How much do you believe you helped the asker?'' and ``How confident are you in your advice (given how much you helped)?''

If a participant was unsure on how to give advice to the question, they could press a button labeled ``I have no idea what the next step should be''. This would send the participant to a separate survey that collected feedback on how to change the question to assist them in giving advice (e.g., ``Could you have given advice to this question if it had more media (images, videos) as part of the explanation?''). 

In both cases, participants were also asked if they thought an expert in the domain could answer the question quickly, even if they could not, as a sanity check on whether the questions we extracted indeed represented potential nanoquestions.


A limitation of this protocol is that the original question askers did not rate the helpfulness of the answers, but rather participants self-rated how helpful their answers were. However, we were comfortable with this trade-off as it enabled testing more questions with more people than otherwise possible. Additionally, the results show no evidence that participants were ``over-rating'' the helpfulness of their answers (participants often said they didn't know the next step, or that their answers were not helpful). 
Finally, the 60-second visual alert may have pressured participants or been distracting. However, none mentioned the alert in the open-ended feedback field.

\subsection{Post-Study Questionnaire}
After completing all 20 questions, participants were given a final survey that collected information on the difficulty in providing advice via text and audio and their preference between modalities (\cref{sec:rq2result}). 
Participants then rated how various changes to the questions they saw might facilitate providing answers (\cref{sec:rq3result}). 
\section{Results}
\label{sec:results}
We verified there was no significant difference between the Word and Fusion groups for: 
overall answer times (t=1.35, p=.177, df=538.8), 
modality preference ($\chi^2$(2, N=28)=0.19, p=.979), 
and helpfulness (Mann-Whitney U: U=10749.5, p=.807).
We did observe a difference between Word and Fusion answer times without ``I don't know'' answers (t=2.77, p=.006, df=247.3; Fusion median = 48.6, mean = 52.42 seconds; Word median = 55.65, mean = 61.57 seconds), which we attribute to the shorter length for Fusion nanoquestions used in the study (Fusion body median = 42, mean = 49.067 words, title median = 6.5, mean = 8.3 words vs. Word body median = 59.5, mean = 63.1 words, title median = 7.5, mean = 8.667 words).
%
In what follows, we pair quantitative results with qualitative feedback to answer our three research questions.

\subsection{RQ1: Can Helpers Give Advice Quickly And Give Advice That They Believe Is Helpful?}
\mysec{Participant Answering Efficiency} 
Participants were able to provide an answer 52\% (291/560) of the time. For these tasks (where an answer was given), the median time for completion was 53.3 seconds (mean = 57.0). For the 48\% of questions that were not answered, the median time for the participant to indicate they did not have the ability to answer a question was 27.6 seconds.

Overall, the median ``read time'' (time from question reveal to clicking the text area or audio record button) for answered questions was 27.6 seconds (mean = 33.6) while the median ``answer time'' (time from starting to answer until submission) was 18.6 seconds (mean = 23.4). Read time represented over half (58.9\%) of the time spent for participants during the tasks, with the remaining 41.1\% of time spent typing or speaking advice to the question (\cref{fig:timeline-teaser}).

These results verify our assumption that many questions could be answered in less than a minute if directly routed to an appropriate and available expert, rather than passively posting online. 
This efficiency could enable users to get unstuck quickly and continue their work, showing potential for integrating nanoquestions into existing workflows of systems for getting quick answers from experts (\cref{sec:rw}).
Only 52\% of the nanoquestions were answered, but this implies 10.4\% of online questions could get quick help as nanoquestions, since they were 20\% of our sample. Moreover, participants thought 91\% of the questions they saw \emph{could} (rated ``Possibly (3)'' or higher) be answered quickly by an expert, even if they could not. This suggests that 18.2\% of forum questions may be nanoquestions. For example, the Fusion 360 sub-forum ``Stuck on a workflow? Have a tricky question about a Fusion 360 feature?'' has over 292,000 posts as of writing. While there may be some duplicates or non-question posts, if these percentages hold there are likely several thousands of nanoquestions on this forum alone. 

However, finding available and relevant helpers is still a critical challenge to getting helpful advice quickly. Previously, it was uncertain if the reason these forum questions have response times of several hours or even days is due to the question being difficult to answer or due to the unavailability of experts to answer the question. Our results motivate the design of systems for quick help by confirming that getting these questions to forum experts directly affords answers very quickly. We discuss further implications of the challenge of expert help availability in \cref{sec:disc}.

\begin{figure}[hb]
\includegraphics[width=.9\columnwidth]{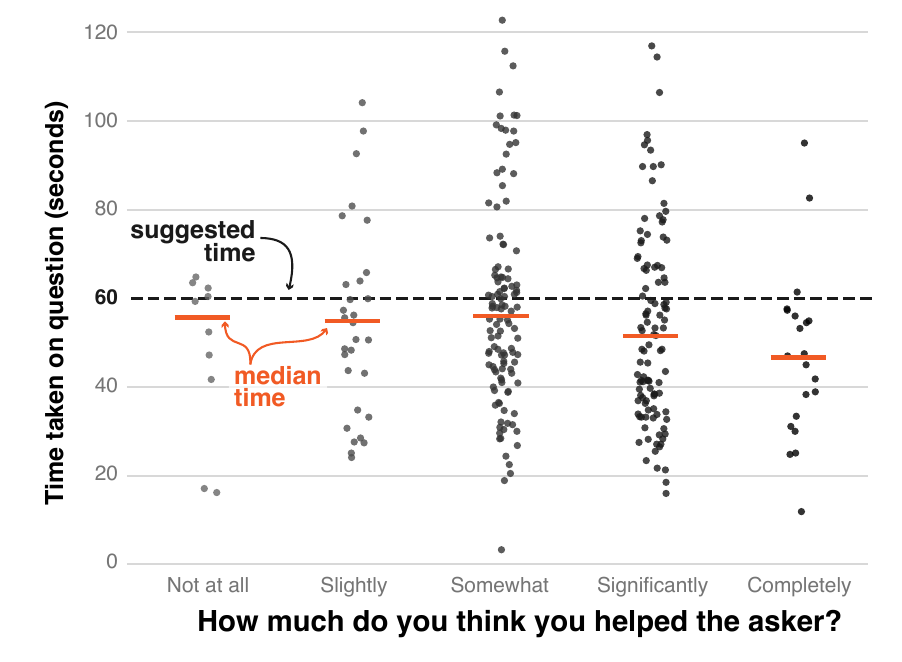}
  \caption{Participants self-rating of much they believed they helped the learner vs time taken on the question. The dotted line represents the 60-second mark. The median time taken on each question for all ratings of helpfulness were below this time.}
  \Description{Participants self-rating of much they believed they helped the learner vs time taken on the question. The dotted line represents the 60-second mark. The median time taken on each question for all ratings of helpfulness were below this time.}
  \label{fig:helpvstime}
\end{figure}

\mysec{Participant-Rated Helpfulness}
After participants gave their answer to each question, they were asked to rate how much they thought they helped the learner on a 5-point Likert scale from ``Not at all (1)'' to ``Completely (5)'' (\cref{fig:helpvstime}). Participants self-rated their help with a median of 3.0 (``Somewhat'', mean = 3.344). 

A one-way ANOVA test with Greenhouse–Geisser correction did not find a significant difference on the amount of time spent and amount of help provided (All- F(4,286) = 1.819, p = .246; Fusion- F(4,140) = 2.072, p = .248; Word- F(4,141) =  1.383, p = .449).
This suggests experts can give helpful advice to learners, even in a very short amount of time. ``Completely'' had the lowest median total time at 47 seconds (mean = 47.01) and ``Somewhat'' had the highest median total time at 56.20 seconds (mean = 59.58).


\subsection{RQ2: Do Helpers Have a Modality Preference When Answering Quickly?}
\label{sec:rq2result}
\begin{figure}[ht]
\includegraphics[width=.9\columnwidth]{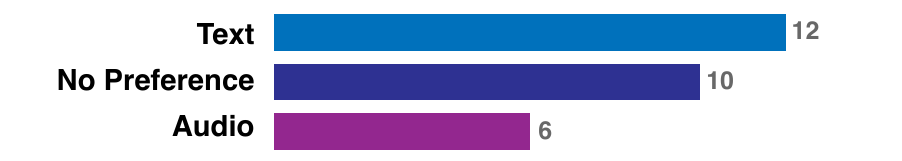}
  \caption{Participants' preferred answer modality: 12 preferred giving advice via text, 10 had no preference, 6 preferred audio.}
  \Description{Participants' preferred answer modality: 12 preferred giving advice via text, 10 had no preference, 6 preferred audio.}
  \label{fig:prefmodal}
\end{figure}

\begin{figure}[ht]
\includegraphics[width=.9\columnwidth]{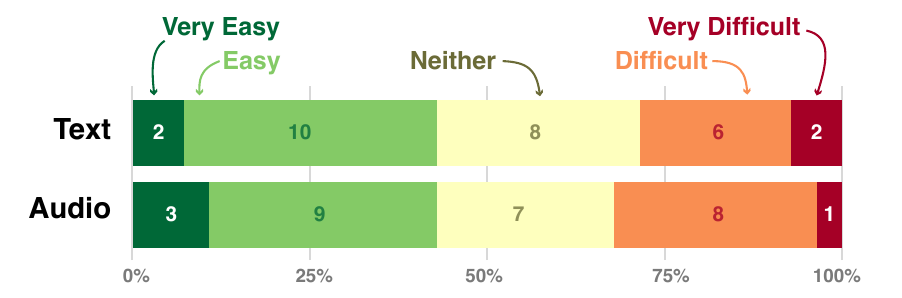}
  \caption{Participants rated the difficulty of using each modality in the study from Very Easy to Very Difficult. The number of participants for each rating is listed in that section of the plot.}
  \Description{Participants rated the difficulty of using each modality in the study from Very Easy to Very Difficult. The number of participants for each rating is listed in that section of the plot.}
  \label{fig:diffmodal}
\end{figure}

\mysec{Modality Preferences}
Participants had varying preferences for answer modality. \cref{fig:prefmodal} shows that while the majority of participants preferred giving answers via text (12), many participants had no preference (10), while some (6) preferred audio. 
When rating how difficult answering was with each modality on a Likert scale from ``Very difficult (1)'' to ``Very easy (5)'', there were mixed results (\cref{fig:diffmodal}). For both text and audio, the median rating was ``Neither easy nor difficult (3)'' (text mean = 3.143; audio mean = 3.179). A Wilcoxon signed-rank test did not find a significant difference for Fusion~360 (W = 18.0, p = .587) or Word (W = 34.5, p = .715).





\mysec{Benefits and Drawbacks of Different Modalities}
Participants were asked to weigh the benefits and drawbacks of quickly providing answers in text and audio. When asked which modality they preferred, several participants said, \emph{``it depends''} (P3, 12, 14, 27). For example, P12 said if they needed to \emph{``give a lot of detail''} they would use audio, otherwise text would be their preference. 

Participants discussed that the ease of answering the question and the speed at which they could provide a satisfying answer was why they preferred one modality over the other (P1, 6, 9, 11, 15, 27). 
P9 said that they \emph{``felt more comfortable with typing since they think faster when writing rather than saying it out loud''}. 
One participant thought that you could  \emph{``go into more detail''} with text, which allowed a more  \emph{``satisfying answer''} (P15). 
However, 
P1 thought it was \emph{``easier to describe the answers verbally, rather than having to type everything out''}. 
P27 said they felt \emph{``audio was WAY faster than text''}, due to the time pressure of answering quickly. 


Participants also thought about going beyond text and audio as a way to provide answers to learners (P5, 14, 16, 18, 21). P18 said it is \emph{``easier to give a considerate response if it is possible to include screenshots to illustrate the procedure''}. P21 agreed, noting that screen sharing and making a video reply to the learner would better demonstrate the process over just textual answers. An annotation feature, where the expert can markup images or videos sent to them, was also seen as potentially useful by some participants (P14, 16).

Participants considered correctness when weighing text against audio (P2, 4, 7, 8, 10, 13). P2 said time pressure made it difficult to \emph{``ensure you have the correct spelling and punctuation when giving advice via text''}. However, participants believed that text allowed for easier proofreading and editing. With text, experts could \emph{``go back and edit the answer if something was worded poorly''} (P13). P10 noted they could not \emph{``do the same for speaking''}. Error correction is critical for providing useful advice, as making mistakes in an answer \emph{``may confuse the person who is asking for advice''} (P4). 

These responses suggest that both text and audio have value as response modalities, and both should be supported.




\subsection{RQ3: What Challenges Do Helpers Face When Giving Advice Quickly?}
\label{sec:rq3result}
\mysec{Difficulties Answering Questions}
Some participants \emph{``felt rushed''} (P3) to give quick answers (P3-5, 8, 18, 19). This was a challenge for providing \emph{``thorough and accurate''} responses (P8) since they wanted to try out their advice (P6) or lookup resources online (P19) to verify their answer before sending it to the asker.


Participants also said that more information could be needed to accurately answers questions, which may include \emph{``going back to the poster for clarification''} (P5).
For question clarity, participants suggested modalities beyond audio and text (P2, 4, 5, 11), like access to the application itself to \emph{``live demo with audio voice-over''} and show UI elements being referenced (P2).
P5 said they frequently leverage screencasting to show the learner what to do. 
P11's main challenge was \emph{``not understanding what the asker is referring to}'' and thought images would help.


Participants also faced issues with unfamiliar features (P6, 7, 9, 13, 25).  P6 said that for \emph{``components of Fusion~360''} they do not use, \emph{``it was not possible to give advice''}.
P9 said a \emph{``reference guide''} would be helpful for questions about unfamiliar features, which allows \emph{``quickly looking up a certain term or abbreviation''} to provide what help they could (P9).
Similarly, some participants had trouble answering questions that had to do with unfamiliar operating systems (P1, 5, 7, 10, 14, 18) or geolocations (P15, 19, 26). 




Finally, some questions were qualitative and asked for opinions or guidance. 
While still possibly suitable as nanoquestions, P7 thought these were \emph{``queries to the community that should be seen by a broad audience''} and should be answered by multiple experts.

\mysec{Improving Ease of Giving Advice}
To gain insights on where to focus efforts on improving question formation, we asked participants about four potential question modifications to rate what percentage of questions they saw would have been easier to answer quickly if they had been altered in these ways (see below, and \cref{fig:improvements}):

{
\vspace{0.5em}
\centering
\resizebox{\columnwidth}{!}{%
\begin{tabular}{m{17em}|m{4em}|m{3em}}
& \textbf{Median} & \textbf{Mean} \\ 
\hline
\vskip 1mm
1. Shorter, more concise text                 & \vskip 1mm 42.5\%          & \vskip 1mm 41.42\%       \\
2. More information through text              & 31.5\%          & 41.04\%       \\
3. More information through images/video   & 50\%            & 48.23\%       \\
4. Access to internal/external information & 65\%            & 62.31\%      
\end{tabular}
\vspace{0.5em}
}
}

Most experts believed that the majority of questions could be improved by enabling access to resources related to the questions they were giving advice to. We primarily envisioned nanoquestions to be answered using one's existing knowledge---but perhaps a dedicated ``nanoquestion answering interface'' could be designed that also provides quick-access to relevant resources.

\begin{figure}[ht]
\includegraphics[width=.9\columnwidth]{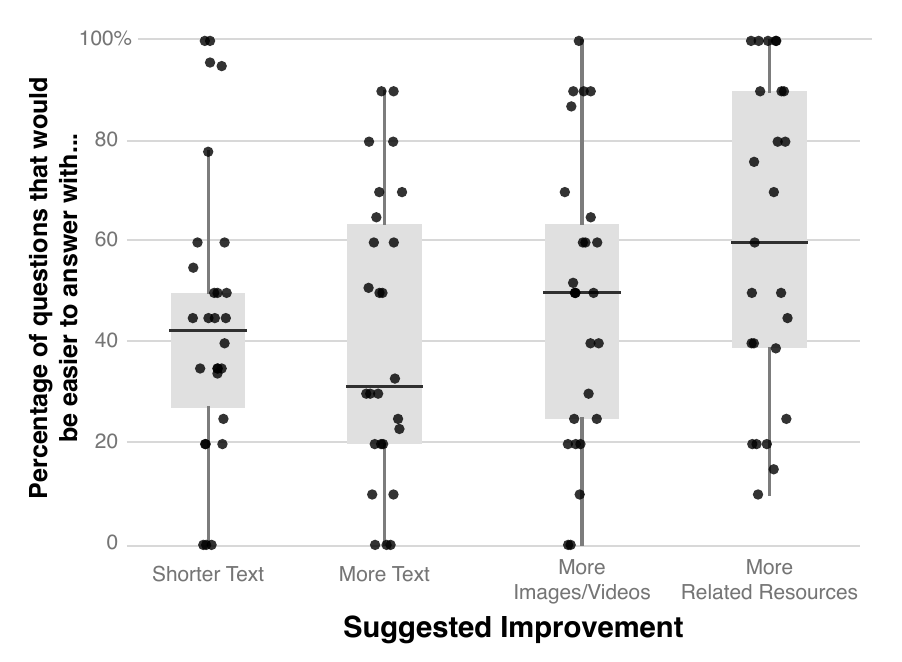}
  \caption{At the end, participants were asked ``of the questions you saw today, what percentage would be significantly easier to provide advice to if the question was modified with: shorter, more concise text; more information through text; more information through images or videos; and access to internal or external information''. Participants rated each change with a percentage.}
  \Description{At the end, participants were asked ``of the questions you saw today, what percentage would be significantly easier to provide advice to if the question was modified with: shorter, more concise text; more information through text; more information through images or videos; and access to internal or external information''. Participants rated each change with a percentage.}
  \label{fig:improvements}
\end{figure}

\mysec{Forming Better Nanoquestions}
Participants suggested how to improve questions for quick advice, addressing many of the difficulties they had during the study.


Participants said images and videos improve a learner's question (P7, 8, 11, 13–15, 17, 19, 24, 28) and \emph{``make it easier to diagnose the issue''} (P7). Images help experts understand \emph{``what the user is seeing''} (P24) and \emph{``what the learner is doing wrong''} (P15).
Participants also wanted access to the content referred to by the question. For example, P5 wanted the \emph{``Fusion 360 model that the user had done''}, since images alone \emph{``are rarely much good at identifying''} the issue, but \emph{``the model tells all''}. 

Many participants offered feedback for improving the content of the question (P2-6, 10, 18). P2 wanted \emph{``better summaries''}, as the question title did not explain the issue. More information, such as the context of \emph{``what the learner is trying to do''} instead of just the problem the learner is trying to solve, is needed (P4).

Though originally selected as nanoquestions, some questions were too complex, which made it difficult to comprehend the question and give an answer (P3, 5, 14, 18). P3 wanted \emph{``shorter questions''}. P10 said this would make questions \emph{``easier to read and answer''} quickly. However, some questions involved designs that were too complex to provide advice quickly, as they \emph{``take time to analyze''} (P5). So, summarizing or decomposing a question may not be enough. The type of answer needed for a question also determined ease of providing quick advice, with questions requiring \emph{``writing out a script or code''} taking more time to answer (P14). 

\section{Discussion}
\label{sec:disc}
\mysec{From Nanoquestions to Micro-Mentorship}
Our participants noted that for some questions further interaction with the learner was needed (e.g., asking follow-up questions for clarification, needing more time to adequately answer the question, sharing screens with the learner to explain a concept). We believe that nanoquestions could be the precursor for opening quick mentorship sessions in a tool such as MicroMentor \cite{NikhitaMicromentor2020}. Nanoquestions can open the dialog between learner and expert, and if needed, the dialog can be extended into a longer session where back-and-forth communication between learner and expert can be supported. 
Finally, sequential nanoquestions might even provide a form of distributed micro-mentorship, 
where many experts can quickly provide help and new experts can jump in when available and their expertise is needed.

\mysec{Nanoquestion Engineering}
We extend previous work on question formation for expert help (\cref{sec:rw}), similar to prompt engineering for LLMs \cite{white2023prompt}. 
We found that human experts can also be better served through certain strategies for asking questions. 
Our results confirm that experts need multimedia and other context to help askers. 
However, there are specific design needs beyond this context for nanoquestions that aim for quick answers. 

There must be a balance between how quickly an expert can read and understand the issue facing the learner. While extra context can help, too much or irrelevant content might prevent the expert providing quick help. 
Approaches like Winder~\cite{KimWinder2021} could better capture the issue for experts through multimodal comments.
For feedback-based questions, multiple experts could give diverse insights quickly to provide consensus on an issue asynchronously. 

Experts said additional modalities would assist in giving quick advice. Experts wanted to draw on the question's images to quickly point out relevant UI elements.
There should also be minimal barriers and effort to providing this mentorship, based on the kind of question being asked.
Some experts enjoy helping others, but do not want to feel like providing help ``is a quiz show'' (P18). Care must be taken to make experts feel appreciated and not like machines, which might include generating ``niceties'' (e.g., friendly greetings). 

\mysec{Limitations and Future Work} This study explored the existence of nanoquestions on online forums, but there are other challenges in the workflow of getting and giving quick help. Critically, finding available experts to answer these questions, which some prior work in Section \ref{sec:rw} addresses. However, our goal was to show that there are indeed questions that could get quick help from experts if they were identified and notified.

Finding available experts is a future direction for quick help and learning research, as availability may affect response time. We believe actively showing experts relevant questions, as done in our study and related work, is more effective than passively posting questions online. For example, in organizations where training and mentoring is valued, helping colleagues may be rewarded or part of job responsibilities. Prior mentoring systems like MicroMentor could also route nanoquestions to expert colleagues for quick help.

LLM-powered AI-assistants like ChatGPT \cite{openai-chatgpt}, Bing Chat \cite{microsoft-bing-chat}, or Google Gemini \cite{google-bard} may help bridge the gap between asker and helper when relevant experts are unavailable. These assistants should be fine-tuned to the nanoquestion's domain, which includes previous questions, documentation, and internal company knowledge and practices. AI-assistants might also help learners refine questions before seeking help, or be an expert's reference guide (\cref{sec:rq3result}).

Nanoquestions have a lower barrier to giving help as they are quick to address and do not require synchronous help sessions, which may entice more experts to provide help. Now that we know nanoquestions exist and are a non-trivial portion of questions being asked online, we can surface this interaction to provide \emph{nanomentorship} to co-workers and new employees. However, further research is needed on how to best find relevant experts while respecting their availability and goals as a helper.
\section{Conclusion}
\label{sec:concl}
This paper explored nanomentorship by identifying potential nanoquestions online that could be answered in less than 60 seconds. We asked 28 online helpers to give quick advice to these questions and provide feedback on their experience. We found most nanoquestions could be answered quickly to the expert's own satisfaction. 
Our findings suggest nanoquestions have potential to empower experts and users with quick help, and provides design implications for future quick help systems that deliver nanoquestions to relevant experts to get users unstuck quickly.





\bibliographystyle{ACM-Reference-Format}
\bibliography{template}
\balance










\end{document}